\documentclass[10pt,twocolumn,aps,pra,amsmath,amssymb,showpacs]{revtex4-1}
\usepackage{bm}
\usepackage{mathrsfs}
\usepackage{graphicx}
\usepackage[usenames,dvipsnames,svgnames,table]{xcolor}
\usepackage[unicode=true,
            pdfusetitle, 
            bookmarks=true,
            bookmarksnumbered=false,
            bookmarksopen=false,
            breaklinks=true,
            pdfborder={0 0 0},
            backref=false,
            colorlinks=true]{hyperref}
\hypersetup{linkcolor=NavyBlue,urlcolor=NavyBlue,citecolor=NavyBlue}

\usepackage{amsthm}

\providecommand{\propositionname}{Proposition}

\usepackage{tikz-cd}

\newcommand{\cH}{\mathcal{H}}

\begin{document}

\title{Entanglement revival can occur only when the system-environment state is not a Markov state}
\author{Iman Sargolzahi}
\email{sargolzahi@neyshabur.ac.ir, sargolzahi@gmail.com}
\affiliation{Department of Physics, University of Neyshabur, Neyshabur, Iran}

\begin{abstract}
Markov states have been defined for tripartite quantum systems. In this paper, we generalize the definition of the Markov states to arbitrary multipartite case and find the general structure of  an important subset of them, which we will call \textit{strong Markov} states. In addition, we focus on an important property of the Markov states: If the initial state of the whole system-environment is a Markov state, then each localized dynamics of the whole system-environment reduces to a localized subdynamics of the system. This provides us a necessary condition for entanglement revival in an open quantum system: Entanglement revival can occur only when the system-environment state is not a Markov state. 
To illustrate (a part of) our results, we consider the case that the environment is modeled as classical. In this case, though the correlation between the system and the environment remains classical during the evolution, the  change of the state of the system-environment, from its initial Markov state to a state which is not a Markov one, leads to  the entanglement revival in the system. This shows that the non-Markovianity of a state is not equivalent to the existence of non-classical correlation in it, in general. 

\end{abstract}


\maketitle
\section{Introduction}

A famous and important relation in quantum information theory is the \textit{strong subadditivity relation}, i.e., for each tripartite quantum state $\rho_{ABE}$, the following inequality holds:
\begin{equation}
\label{eq:1000}
\begin{aligned}
S(\rho_{AB})+ S(\rho_{BE})-S(\rho_{ABE})-S(\rho_{B}) \geq 0, 
\end{aligned}
\end{equation}
where $\rho_{AB}=\mathrm{Tr_{E}}(\rho_{ABE})$, $\rho_{BE}=\mathrm{Tr_{A}}(\rho_{ABE})$  
and $\rho_{B}=\mathrm{Tr_{AE}}(\rho_{ABE})$ are the reduced states
 and $S(\rho)\equiv - \mathrm{Tr}(\rho log \rho)$ is the von Neumann entropy \cite{13}.

Markov states have been defined, in Ref. \cite{10}, as tripartite quantum states which satisfy the strong subadditivity relation  with equality. Recently, Markov states  have been applied in studying the dynamics of open quantum systems  \cite{11,12}. 

In this paper, we generalize the definition of the Markov states to arbitrary multipartite case. Our definitions will be given in two forms, a \textit{weak} one and a \textit{strong} one. The strong form is more restricted than the weak form. In addition, we find the general structure of the \textit{strong Markov} (SM) states. 

The above results will be given during our study of the role of the Markov states in  entanglement dynamics of  open quantum systems. This will help us to give the definitions, and so the subsequent related results, such that they have clear physical meanings and applications.

 The dynamics of the entanglement in open quantum systems, both
in bipartite and multipartite cases, has been studied widely 
 \cite{1}. Entanglement may decrease or even experience revivals during the interaction of the system with the environment \cite{2}. 

Consider a bipartite system $S=AB$ such that each part interacts with its local environment. One may expect that in this case only entanglement decrease (sudden death) will occur, since entanglement does not increase under local operations. But, interestingly, it has been shown, both theoretically and experimentally, that the entanglement revival can occur under such circumstances (see,  e.g.,  \cite{23, 24, 8}). More unexpectedly, entanglement revival can occur even when the environment is classical \cite{3, 4, 5, 6, 7, 8}.

We question when entanglement revival can occur, under local interactions, and find a necessary condition for this phenomenon. 
 We show that the entanglement revival (increase) can occur only when the whole state of the  system-environment is not Markov state. This necessary condition is valid for both bipartite and multipartite cases.
 
 Usually, the initial state of the system-environment is chosen factorized, which is a Markov state. In addition, the  dynamics of the system-environment is given by local unitary operators. So, the entanglement of the system $S$, initially, starts to decrease and if, e.g., at time $t$ the  entanglement of the system starts to revive (increase), then we conclude that $\rho_{SE}(t)$, the state of the system-environment at time $t$, is not a Markov state.

 In the next section, we consider the simplest case: When only the part $B$ of our bipartite system $S=AB$ interacts with the environment $E$. We recall the original definition of the Markov states from Ref. \cite{10}, and we will see that the entanglement revival can occur only  when $\rho_{ABE}$, the whole state of the system-environment, is not a Markov state.
 
 As stated before, the original definition of the Markov states in Ref. \cite{10} is for the tripartite case.  
  In Sect.~\ref{sec:individual environment}, we extend the definition of the Markov states to the quadripartite case. We give two definitions, a weak one and, a more restricted form, a strong one.  We
  generalize the result of  Sect.~\ref{sec:part B},  about entanglement revival, to the case that each part of the system $S=AB$ interacts  with its local  environment. This will be done, using the weak definition. In addition, we
 give our first main result as Theorem 3.  This theorem gives us the general structure of the quadripartite strong Markov (SM) states. 
  
  In Sect.~\ref{sec:multi},   we generalize our results  to arbitrary multipartite case. We give our second main result as Theorems 4 and 5. In these theorems, we  find the general structure of the SM states for arbitrary multipartite case.  
  In addition, we show that, as the previous sections,   
    if the initial state of the system-environment is a weak Markov (WM) state, then each localized dynamics of the whole system-environment reduces to a localized subdynamics of the system.
    Therefore, also for the multipartite case, entanglement revival can occur only when  the initial state of the system-environment is not a WM state.
  
To illustrate (a part of) our results, we consider the case that the environment is classical in Sect.~\ref{sec:classical}. Though the correlation between the system and the environment remains classical during the evolution, entanglement revival can occur. So, the whole state of the  system-environment changes from its initial Markov state to a state which is not a Markov one. This implies that non-Markovianity of the whole state of the  system-environment is not equivalent to the existence of non-classical correlation between the system and the environment.

Finally,    we end our paper in Sect.~\ref{sec:summary}, with a summary of our results.

\section{When only the part $B$ interacts with the environment}\label{sec:part B}


Consider a bipartite quantum system $S=AB$, such that the part $A$ is isolated from the environment, and only the part $B$ interacts with the environment $E$. So, we have
\begin{equation}
\label{eq:101}
\begin{aligned}
\rho_{ABE}^{\prime}=id_{A}\otimes Ad_{U_{BE}}(\rho_{ABE})  \quad \qquad  \\
\equiv I_{A}\otimes U_{BE}\rho_{ABE}I_{A}\otimes U_{BE}^{\dagger},  
\end{aligned}
\end{equation}
where $\rho_{ABE}$ ($\rho_{ABE}^{\prime}$) is the initial (final) state of the system- environment, $id_{A}$ ($I_{A}$) is the identity map (operator) on the part $A$ and $ U_{BE}$ is a unitary operator on the both $B$ and $E$.

Now, assume that we have
\begin{equation}
\label{eq:102}
\begin{aligned}
\rho_{ABE}=id_{A}\otimes \Lambda_{B}(\rho_{AB}),  
\end{aligned}
\end{equation}
where $\rho_{AB}=\mathrm{Tr_{E}}(\rho_{ABE})$ is the initial state of the system $S=AB$ and 
 $\Lambda_{B}$ is a completely positive (CP) map from $B$ to $BE$. [A completely positive map, on a state $\rho$ , is a map which can be written as 
$\sum_{i} K_{i}\rho K_{i}^{\dagger}$, where $K_{i}$ are linear operators such that 
$\sum_{i} K_{i}^{\dagger} K_{i}=I$  ($I$ is the identity operator) \cite{13}.] 
If Eq. (\ref{eq:102}) holds, the tripartite state $\rho_{ABE}$ is called a Markov state \cite{10}:

\textbf{Definition 1.} \textit{A tripartite state $\rho_{ABE}$ is called a Markov state, if it can be written as Eq. (\ref{eq:102}).}
 
 It can be shown that, for a tripartite state $\rho_{ABE}$,  the strong subadditivity inequality, i.e.,  Eq. (\ref{eq:1000}), holds  with equality, if and only if,  Eq. (\ref{eq:102}) holds (if and only if,  $\rho_{ABE}$ is a Markov state) \cite{10}.

From Eqs. (\ref{eq:101}) and (\ref{eq:102}), we have
\begin{equation}
\label{eq:103}
\begin{aligned}
\rho_{AB}^{\prime}=\mathrm{Tr_{E}}(\rho_{ABE}^{\prime}) \qquad\qquad\qquad\qquad\quad\quad\qquad   \\
=\mathrm{Tr_{E}}\circ [id_{A}\otimes Ad_{U_{BE}}]\circ [id_{A}\otimes \Lambda_{B}](\rho_{AB}) \\
=id_{A}\circ [\mathrm{Tr_{E}}\circ Ad_{U_{BE}}\circ \Lambda_{B}](\rho_{AB}) \qquad\quad\;\;\\
=id_{A}\circ \mathcal{E}_{B}(\rho_{AB}),\qquad\qquad\qquad\qquad\qquad
\end{aligned}
\end{equation}
where $\rho_{AB}^{\prime}$ is the final state of the system $S=AB$ and $\mathcal{E}_{B}=\mathrm{Tr_{E}}\circ Ad_{U_{BE}}\circ \Lambda_{B}$ is a CP map on the part $B$. $\mathcal{E}_{B}$ is CP since it is a  composition of three CP maps: $\Lambda_{B}$ is CP by assumption, $ Ad_{U_{BE}}$ is obviously CP and the CP-ness of $\mathrm{Tr_{E}}$ can be shown easily \cite{13}.

In other words, if the initial state of the whole system-environment is a Markov state as Eq. (\ref{eq:102}), then each localized dynamics, as  Eq.  (\ref{eq:101}), reduces to a localized subdynamics as Eq.  (\ref{eq:103}). In fact, we have 
the following theorem, which is proven in Ref. \cite{11}. We give it in the form introduced in Ref. \cite{12}, which is appropriate for our purpose in this paper.

\textbf{Theorem 1.} \textit{If, for a tripartite state $\rho_{ABE}$, each \textit{localized} dynamics as }
\begin{equation}
\label{eq:5}
\begin{aligned}
\rho_{ABE}^{\prime}=id_{A}\otimes \mathcal{F}_{BE} (\rho_{ABE})\qquad \qquad\qquad \qquad \quad   \\
=\sum_{j}\left( I_{A}\otimes F_{BE}^{(j)}\right)\,\rho_{ABE}\, \left( I_{A}\otimes F_{BE}^{(j) \dagger}\right),\\
\sum_{j}F_{BE}^{(j) \dagger}F_{BE}^{(j)}=I_{BE},  \qquad \qquad
\end{aligned}
\end{equation}
\textit{ reduces to a localized subdynamics as }
\begin{equation}
\label{eq:6}
\begin{aligned}
\rho_{AB}^{\prime}=id_{A}\otimes \mathcal{E}_{B} (\rho_{AB})\qquad \qquad\qquad \qquad \quad   \\
=\sum_{i}\left( I_{A}\otimes E_{B}^{(i)}\right)\,\rho_{AB}\, \left( I_{A}\otimes E_{B}^{(i)\dagger}\right), \\
\sum_{i} E_{B}^{(i)\dagger}E_{B}^{(i)}=I_{B},\qquad\qquad
\end{aligned}
\end{equation}
\textit{then $\rho_{ABE}$ is a Markov state as Eq. (\ref{eq:102}), and vice versa. In Eq. (\ref{eq:5}),
 $F_{BE}^{(j)}$ are linear operators on $BE$ and, in Eq. (\ref{eq:6}),
  $E_{B}^{(i)}$ are linear operators on $B$.}

The inverse part of Theorem 1 states that if  $\rho_{ABE}$ is a Markov state, then each localized dynamics as  Eq. (\ref{eq:5}) reduces to a localized subdynamics as Eq. (\ref{eq:6}). Let 
$\mathcal{M}$ be  an  entanglement monotone (measure). So, $\mathcal{M}(\rho_{AB}^{\prime})\leq \mathcal{M}(\rho_{AB})$, since entanglement does not increase under local operations as Eq. (\ref{eq:6}) (see, e.g., Ref. \cite{1}).
 Therefore,
 
 \textbf{Corollary 1.} \textit{If for a localized dynamics of the whole system-environment as Eq. (\ref{eq:5}), the entanglement of the system $S=AB$ increases: $\mathcal{M}(\rho_{AB}^{\prime})> \mathcal{M}(\rho_{AB})$, then  we conclude that the initial state of the whole system-environment,  $\rho_{ABE}$, is not a Markov state as Eq. (\ref{eq:102}).}

The following point is also worth noting. Assume that for all $t\in \left[t_{1}, t_{1}^{\prime}\right] $, the entanglement of the system $S=AB$ increases monotonically.
Since we have considered the time evolution of the system-environment as Eq. (\ref{eq:101}),
  the time evolution operator from $t$ to $t_{2}$,  $U_{ABE}(t_{2},t)$, $t< t_{2}$, is also localized as $I_{A}\otimes U_{BE}(t_{2},t)$. Therefore, for each $t\in \left[t_{1}, t_{1}^{\prime}\right) $, the state of the system-environment  $\rho_{ABE}(t)$ is not a Markov state as Eq. (\ref{eq:102}).

Let's end this section with a theorem, proven in Ref. \cite{10}, which gives the general structure of the Markov states, for the tripartite case.

\textbf{Theorem 2.} 
 \textit{A tripartite state $\rho_{ABE}$ is  a Markov state as Eq. (\ref{eq:102}), if and only if,  there exists a decomposition of the Hilbert space of the subsystem $B$, $\cH_{B}$, as $\cH_{B}=\bigoplus_{k}\cH_{b^{L}_{k}}\otimes\cH_{b^{R}_{k}}$  such that}
\begin{equation}
\label{eq:4}
\rho_{ABE}=\bigoplus_{k}\lambda_{k}\:\rho_{Ab^{L}_{k}}\otimes\rho_{b^{R}_{k}E},
\end{equation}
\textit{where $\lbrace \lambda_{k}\rbrace$ is a probability distribution ($\lambda_{k}\geq 0$, $\sum_{k}\lambda_{k}=1$), $\rho_{Ab^{L}_{k}}$ is a state on $\cH_{A}\otimes\cH_{b^{L}_{k}}$ and $\rho_{b^{R}_{k}E}$ is a state on $\cH_{b^{R}_{k}}\otimes\cH_{E}$. ($\cH_{A}$ and $\cH_{E}$ are the Hilbert spaces of $A$ and $E$, respectively.)}

\textbf{Remark 1.}\textit{ Theorem 2 is valid for the case that $\cH_{B}$ is finite dimensional, but $\cH_{A}$ and $\cH_{E}$ can be infinite dimensional. The condition that $\cH_{B}$ is finite dimensional comes from the fact that the proof of Theorem 2 in Ref. \cite{10} is based on a result, proven in Ref. \cite{17}, which is for the finite dimensional case.} 

\section{When each part of the system interacts with its local environment}\label{sec:individual environment}
Now, let's consider the case that the two parts $A$ and $B$ of our bipartite system are separated from each other and each part interacts with its own local environment. Let's denote the local environment of $A$ as $E_{A}$, the local environment of $B$ as $E_{B}$ and the whole state of the system-environment as $\rho_{AE_{A}BE_{B}}$.

First, we generalize the definition of the Markov states to the quadripartite case. The original definition, in Eq.  (\ref{eq:102}), is for the tripartite case.

\textbf{Definition 2.} \textit{We call a quadripartite state $\rho_{AE_{A}BE_{B}}$  a weak Markov (WM) state if there exist CP maps $\Lambda_{A}$, from $A$ to $AE_{A}$, and  $\Lambda_{B}$, from $B$ to $BE_{B}$, such that}
\begin{equation}
\label{eq:104}
\begin{aligned}
\rho_{AE_{A}BE_{B}}=\Lambda_{A}\otimes \Lambda_{B}(\rho_{AB}),  
\end{aligned}
\end{equation}
\textit{where $\rho_{AB}= \mathrm{Tr_{E_{A}E_{B}}}(\rho_{AE_{A}BE_{B}})$.}

If Eq. (\ref{eq:104}) holds, then each localized dynamics as $\mathcal{F}_{AE_{A}}\otimes \mathcal{F}_{BE_{B}}$, for the whole system-environment, reduces to a localized subdynamics as $\mathcal{E}_{A}\otimes \mathcal{E}_{B}$, for the system:
\begin{equation}
\label{eq:105}
\begin{aligned}
\rho_{AB}^{\prime}=\mathrm{Tr_{E_{A}E_{B}}}(\rho_{AE_{A}BE_{B}}^{\prime}) \qquad\qquad\qquad\qquad\qquad\quad\qquad   \\
=\mathrm{Tr_{E_{A}E_{B}}}\circ [\mathcal{F}_{AE_{A}}\otimes \mathcal{F}_{BE_{B}}]\circ [\Lambda_{A}\otimes \Lambda_{B}](\rho_{AB}) \quad\quad\\
=[\mathrm{Tr_{E_{A}}}\circ \mathcal{F}_{AE_{A}}\circ \Lambda_{A}]\circ [\mathrm{Tr_{E_{B}}}\circ \mathcal{F}_{BE_{B}}\circ \Lambda_{B}](\rho_{AB}) \;\;\\
=\mathcal{E}_{A}\circ \mathcal{E}_{B}(\rho_{AB}),\qquad\qquad\qquad\qquad\qquad\qquad\qquad\quad
\end{aligned}
\end{equation}
where $\rho_{AE_{A}BE_{B}}^{\prime}$ ($\rho_{AB}^{\prime}$) is the final state of the system-environment (system). Therefore, $\mathcal{M}(\rho_{AB}^{\prime})\leq \mathcal{M}(\rho_{AB})$. In other words,

\textbf{Corollary 2.} \textit{If for a localized dynamics of the whole system-environment as $\mathcal{F}_{AE_{A}}\otimes \mathcal{F}_{BE_{B}}$, the entanglement of the system $S=AB$ increases: $\mathcal{M}(\rho_{AB}^{\prime})> \mathcal{M}(\rho_{AB})$, then  we conclude that the initial state of the whole system-environment,  $\rho_{AE_{A}BE_{B}}$, is not a WM state as Eq. (\ref{eq:104}).}

Consider a special case that the localized dynamics of the whole system-environment is as $id_{AE_{A}}\otimes \mathcal{F}_{BE_{B}}$. So, from Eq. (\ref{eq:105}), we have
\begin{equation}
\label{eq:106}
\begin{aligned}
\rho_{AB}^{\prime}
=\Phi_{A}\circ \mathcal{E}_{B}(\rho_{AB}),
\end{aligned}
\end{equation}
where $\Phi_{A} =\mathrm{Tr_{E_{A}}} \circ \Lambda_{A}$ is a CP map on $A$. Note that 
$\rho_{A} = \mathrm{Tr_{B}}(\rho_{AB})$ does not change during the evolution.
 So, a natural  requirement, which we may want to add, is that,  for arbitrary CP map $\mathcal{E}_{B}$ on $B$, we must have $\Phi_{A}\circ \mathcal{E}_{B}(\rho_{AB})=id_{A}\circ \mathcal{E}_{B}(\rho_{AB})$. Similarly, for arbitrary CP map $\mathcal{E}_{A}$ on $A$, we must have $\mathcal{E}_{A}\circ \Phi_{B}(\rho_{AB})= \mathcal{E}_{A}\circ id_{B}(\rho_{AB})$, where $\Phi_{B} =\mathrm{Tr_{E_{B}}} \circ \Lambda_{B}$ is a CP map on $B$. 

The above discussion leads us to the following definition:

\textbf{Definition 3.} \textit{We call 
a quadripartite state $\rho_{AE_{A}BE_{B}}$  a strong Markov (SM) state if}

\textit{1. there exist CP maps $\Lambda_{A}$, from $A$ to $AE_{A}$, and  $\Lambda_{B}$, from $B$ to $BE_{B}$, such that
$\rho_{AE_{A}BE_{B}}=\Lambda_{A}\otimes \Lambda_{B}(\rho_{AB})$, and}

\textit{2. for each arbitrary CP maps $\mathcal{E}_{A}$ and $\mathcal{E}_{B}$, we have  $\mathcal{E}_{A}\circ \Phi_{B}(\rho_{AB})= \mathcal{E}_{A}\circ id_{B}(\rho_{AB})$  and $\Phi_{A}\circ \mathcal{E}_{B}(\rho_{AB})=id_{A}\circ \mathcal{E}_{B}(\rho_{AB})$, respectively.}

 In   the following of this section, we will prove our first main result: The general structure of  the  quadripartite SM states. 

\textbf{Theorem 3.} \textit{A quadripartite state $\rho_{AE_{A}BE_{B}}$ is  a SM state, if and only if,  there exist  decompositions of the Hilbert spaces of the subsystems $A$, $\cH_{A}$, and $B$, $\cH_{B}$, as $\cH_{A}=\bigoplus_{j}\cH_{a_{j}^{L}}\otimes\cH_{a_{j}^{R}}$ and $\cH_{B}=\bigoplus_{k}\cH_{b^{L}_{k}}\otimes\cH_{b^{R}_{k}}$, respectively,  such that} 
\begin{equation}
\label{eq:8}
\begin{aligned}
\rho_{AE_{A}BE_{B}}
=\bigoplus_{j, k}\lambda_{jk}\,\rho_{a_{j}^{L}E_{A}}\otimes\rho_{a_{j}^{R}b_{k}^{L}}\otimes\rho_{b_{k}^{R}E_{B}}.
\end{aligned}
\end{equation}
\textit{In Eq.  (\ref{eq:8}),  $\lbrace \lambda_{jk}\rbrace$  is a probability distribution, $\rho_{a_{j}^{L}E_{A}}$ is a state on $\cH_{a_{j}^{L}}\otimes\cH_{E_{A}}$, $\rho_{a_{j}^{R}b_{k}^{L}}$ is a state on $ \cH_{a_{j}^{R}}\otimes\cH_{b_{k}^{L}}$ and $\rho_{b_{k}^{R}E_{B}}$ is a state on $\cH_{b_{k}^{R}}\otimes\cH_{E_{B}}$. ($ \cH_{E_{A}}$ and $ \cH_{E_{B}}$ are the Hilbert spaces of $E_{A}$ and  $E_{B}$, respectively.) }


\textit{Proof.} Showing that the state given in Eq. (\ref{eq:8}) can be written as Eq. (\ref{eq:104}) has been done in Ref. \cite{12}. In addition, using the result of Ref. \cite{12}, showing that the second property in Definition 3 is also fulfilled, by a state as Eq. (\ref{eq:8}), is simple.
 So, in the following, we focus on the reverse: Each quadripartite SM state can be written as   Eq. (\ref{eq:8}). 

First,  
note  that the CP map $\Lambda_{B}$, in Eq. (\ref{eq:104}),  is a map from $B$ to $BE_{B}$.
 To make the input and output spaces the same, we redefine     $\Lambda_{B}$ in the following way: If for an operator $x$ on $B$ we have $\Lambda_{B}(x)=X$, where $X$ is a operator on  $BE_{B}$,  we set $\Lambda_{B}(x\otimes\vert 0_{E_{B}}\rangle\langle 0_{E_{B}}\vert)=X$ where $\vert 0_{E_{B}}\rangle$ is a fixed state in $\cH_{E_{B}}$. This redefinition allows us to write $\Lambda_{B}$ in the following form. One can find an ancillary Hilbert space $\cH_{C_{B}}$, a fixed state $\vert 0_{C_{B}}\rangle\in\cH_{C_{B}}$ and a unitary operator $V_{B}$ on $\cH_{B}\otimes\cH_{E_{B}}\otimes\cH_{C_{B}}$ in such a way that the CP map $\Lambda_{B}$ can be written as \cite{13}:
\begin{equation}
\label{eq:twelve}
\begin{aligned}
\Lambda_{B}(x)=\Lambda_{B}(x\otimes\vert 0_{E_{B}}\rangle\langle 0_{E_{B}}\vert) \qquad\qquad\qquad\quad \;\;\\
=\mathrm{Tr_{C_{B}}}\left(V_{B}\,(x\otimes\vert 0_{E_{B}}\rangle\langle 0_{E_{B}}\vert\otimes\vert 0_{C_{B}}\rangle\langle 0_{C_{B}}\vert)\,V^{\dagger}_{B}\right).
\end{aligned}
\end{equation}

Also, note that for the CP map $\Phi_{B}$,
 on the $B$, we have 
\begin{equation}
\label{eq:fourteen}
\begin{aligned}
\Phi_{B}(x)
=\mathrm{Tr_{E_{B}C_{B}}}\left(V_{B}\,(x \otimes\vert 0_{E_{B}}\rangle\langle 0_{E_{B}}\vert\otimes\vert 0_{C_{B}}\rangle\langle 0_{C_{B}}\vert)\,V_{B}^{\dagger}\right),
\end{aligned}
\end{equation}
with the unitary $V_{B}$ introduced in Eq. (\ref{eq:twelve}). Similar results can be driven for the CP maps $\Lambda_{A}$ and $\Phi_{A}$.

Second, from Eq. (\ref{eq:104}), we have
\begin{equation*}
\label{eq:107}
 \Phi_{A}\otimes \Phi_{B}(\rho_{AB})=\rho_{AB}.
 \end{equation*}
  So, using the property 2 in Definition 3, we can rewrite the above equation as
\begin{equation}
\label{eq:108}
 id_{A}\otimes \Phi_{B}(\rho_{AB})=\rho_{AB}.
 \end{equation}   
Now, from the proof of Theorem 2 in Ref. \cite{10}, we know that if Eq. (\ref{eq:108}) holds, then 
there exists a decomposition of the $\cH_{B}$ as $\cH_{B}=\bigoplus_{k}\cH_{b^{L}_{k}}\otimes\cH_{b^{R}_{k}}$  such that:

1. $\rho_{AB}$ can be decomposed as
\begin{equation}
\label{eq:109}
\rho_{AB}=\bigoplus_{k}q_{k}\:\rho_{Ab^{L}_{k}}\otimes\rho_{b^{R}_{k}},
\end{equation}
where $\lbrace q_{k}\rbrace$ is a probability distribution,  $\rho_{Ab^{L}_{k}}$ is a state on $\cH_{A}\otimes\cH_{b^{L}_{k}}$ and $\rho_{b^{R}_{k}}$ is a state on $\cH_{b^{R}_{k}}$, and 

2. the unitary operator $V_{B}$, in Eq. (\ref{eq:fourteen}), is as 
\begin{equation}
\label{eq:sixteen}
\begin{aligned}
V_{B}=\bigoplus_{k}I_{b^{L}_{k}}\otimes V_{b^{R}_{k}E_{B}C_{B}},
\end{aligned}
\end{equation}
where $I_{b^{L}_{k}}$ is the identity operator on $\cH_{b^{L}_{k}}$ and $V_{b^{R}_{k}E_{B}C_{B}}$ is a unitary operator on $\cH_{b^{R}_{k}}\otimes\cH_{E_{B}}\otimes\cH_{C_{B}}$.

 Also note that, since during the proof of Theorem 2 in Ref. \cite{10} a result of Ref. \cite{17} has been used, $\cH_{B}$ is finite dimensional.

Similarly, starting from $ \Phi_{A}\otimes id_{B}(\rho_{AB})=\rho_{AB}$, it can be shown that 
there exists a decomposition of the finite dimensional Hilbert space $\cH_{A}$ as $\cH_{A}=\bigoplus_{j}\cH_{a^{L}_{j}}\otimes\cH_{a^{R}_{j}}$  such that:

1. $\rho_{AB}$ can be decomposed as
\begin{equation}
\label{eq:110}
\rho_{AB}=\bigoplus_{j} p_{j}\:\rho_{a^{L}_{j}}\otimes\rho_{a^{R}_{j}B},
\end{equation}
where $\lbrace p_{j}\rbrace$ is a probability distribution,   $\rho_{a^{L}_{j}}$ is a state on $\cH_{a^{L}_{j}}$ and  $\rho_{a^{R}_{j}B}$ is a state on $\cH_{a^{R}_{j}}\otimes \cH_{B}$, and

2. the unitary operator $V_{A}$  is as 
\begin{equation}
\label{eq:110-1}
\begin{aligned}
V_{A}=\bigoplus_{j} V_{a^{L}_{j}E_{A}C_{A}}\otimes I_{a^{R}_{j}},
\end{aligned}
\end{equation}
where $I_{a^{R}_{j}}$ is the identity operator on $\cH_{a^{R}_{j}}$ and $V_{a^{L}_{j}E_{A}C_{A}}$ is a unitary operator on $\cH_{a^{L}_{j}}\otimes\cH_{E_{A}}\otimes\cH_{C_{A}}$.

Third, consider the projection
\begin{equation}
\label{eq:thirty six}
\begin{aligned}
\Pi_{j}\equiv\Pi_{A_{j}}\otimes I_{B}\qquad\qquad\\
=(\Pi_{a_{j}^{L}}\otimes\Pi_{a_{j}^{R}})\otimes I_{B},
\end{aligned}
\end{equation}
where $\Pi_{A_{j}}$,  $\Pi_{a_{j}^{L}}$ and $\Pi_{a_{j}^{R}}$ are the projectors onto $\cH_{A_{j}}=\cH_{a_{j}^{L}}\otimes\cH_{a_{j}^{R}} $,  $\cH_{a_{j}^{L}}$ and $\cH_{a_{j}^{R}}$, respectively. So, from Eqs. (\ref{eq:109}) and (\ref{eq:110}), we have
\begin{equation}
\label{eq:111}
\Pi_{j}\rho_{AB}\Pi_{j}= p_{j}\:\rho_{a^{L}_{j}}\otimes\rho_{a^{R}_{j}B}=\bigoplus_{k}q_{k}\:\sigma_{A_{j}b^{L}_{k}}\otimes\rho_{b^{R}_{k}},
\end{equation}
where $\sigma_{A_{j}b_{k}^{L}}=\bar{\Pi}_{j}\,\rho_{Ab_{k}^{L}}\,\bar{\Pi}_{j}$ and $\bar{\Pi}_{j}\equiv\Pi_{A_{j}}\otimes\Pi_{b_{k}^{L}}$ (where $\Pi_{b_{k}^{L}}$ is the projection onto $\cH_{b_{k}^{L}}$). $\sigma_{A_{j}b_{k}^{L}}$ is a positive operator on $\cH_{A_{j}}\otimes\cH_{b_{k}^{L}}$. Let $p^{\prime}_{jk}=\mathrm{Tr}(\sigma_{A_{j}b_{k}^{L}})$; so $0\leq p^{\prime}_{jk}\leq 1$. Now if $p^{\prime}_{jk}>0$, we define
\begin{equation*}
\rho_{A_{j}b_{k}^{L}}=\dfrac{\sigma_{A_{j}b_{k}^{L}}}{p^{\prime}_{jk}} ,
\end{equation*}
otherwise, if $p^{\prime}_{jk}=0$, we define $\rho_{A_{j}b_{k}^{L}}$ arbitrarily.
So, Eq.~(\ref{eq:111}) can be rewritten as
\begin{equation*}
p_{j}\,\rho_{a_{j}^{L}}\otimes\rho_{a_{j}^{R}B}
=\bigoplus_{k}q_{k}p^{\prime}_{jk}\,\rho_{A_{j}b_{k}^{L}}\otimes\rho_{b_{k}^{R}}.\,
\end{equation*}
Tracing from both sides, with respect to $a_{j}^{L}$, we get
\begin{equation}
\label{eq:fifty}
p_{j}\rho_{a_{j}^{R}B}
=\bigoplus_{k}\lambda_{jk}\,\rho_{a_{j}^{R}b_{k}^{L}}\otimes\rho_{b_{k}^{R}},
\end{equation}
where $\rho_{a_{j}^{R}b_{k}^{L}}=\mathrm{Tr_{a_{j}^{L}}}(\rho_{A_{j}b_{k}^{L}})$ 
 and $\lambda_{jk}=q_{k}p^{\prime}_{jk}$. Therefore, Eq. (\ref{eq:110}) can be rewritten as
  \begin{equation}
\label{eq:112}
\rho_{AB}=\bigoplus_{j,k} \lambda_{jk}\:\rho_{a^{L}_{j}}\otimes\rho_{a_{j}^{R}b_{k}^{L}}\otimes\rho_{b_{k}^{R}}.
\end{equation}

Fourth, combining Eqs. (\ref{eq:twelve}) and (\ref{eq:sixteen})  gives us  
\begin{equation}
\label{eq:seventeen}
\begin{aligned}
\Lambda_{B}=\bigoplus_{k}id_{b_{k}^{L}}\otimes \Lambda_{b_{k}^{R}},
\end{aligned}
\end{equation}
where $id_{b_{k}^{L}}$ is the identity map on $b_{k}^{L}$ and $\Lambda_{b_{k}^{R}}$ is a CP map from $b_{k}^{R}$ to $b_{k}^{R}E_{B}$. Similarly, we have
\begin{equation}
\label{eq:113}
\begin{aligned}
\Lambda_{A}=\bigoplus_{j}\Lambda_{a_{j}^{L}}\otimes id_{a_{j}^{R}},
\end{aligned}
\end{equation}
where $\Lambda_{a_{j}^{L}}$ is a CP map from $a_{j}^{L}$ to $a_{j}^{L}E_{A}$.

Finally, using Eqs. (\ref{eq:104}), (\ref{eq:112}), (\ref{eq:seventeen}) and (\ref{eq:113}), we achieve Eq. (\ref{eq:8}), which completes the proof.
$\qquad\qquad\quad\qquad\square$

\textbf{Remark 2.} \textit{As Theorem 2, Theorem 3 is valid for the case that $\cH_{A}$ and $\cH_{B}$ are finite dimensional, but $\cH_{E_{A}}$ and $\cH_{E_{B}}$ can be infinite dimensional. In other words, the system $S=AB$ is finite dimensional but the environments  $E_{A}$ and $E_{B}$ can be infinite dimensional.}

Note that, during the proof of Theorem 3, we only require that
$\Phi_{A}\otimes \Phi_{B}(\rho_{AB})=id_{A}\otimes \Phi_{B}(\rho_{AB})=\Phi_{A}\otimes id_{B}(\rho_{AB})=\rho_{AB}$. So, we can give the definition of the SM states in a less restricted form, as follows:

\textbf{Definition $3^{\prime}$.} \textit{We call 
a quadripartite state $\rho_{AE_{A}BE_{B}}$  a strong Markov (SM) state if}

\textit{1. there exist CP maps $\Lambda_{A}$, from $A$ to $AE_{A}$, and  $\Lambda_{B}$, from $B$ to $BE_{B}$, such that
$\rho_{AE_{A}BE_{B}}=\Lambda_{A}\otimes \Lambda_{B}(\rho_{AB})$, and}

\textit{2. $\Phi_{A}\otimes \Phi_{B}(\rho_{AB})=id_{A}\otimes \Phi_{B}(\rho_{AB})=\Phi_{A}\otimes id_{B}(\rho_{AB})=\rho_{AB}$.}

For each quadripartite state $\rho_{AE_{A}BE_{B}}$, which possesses  the two properties in  Definition $3^{\prime}$, Theorem 3 is valid. Now, using Eqs. (\ref{eq:112}), (\ref{eq:seventeen}) and (\ref{eq:113}), it can be shown simply that the property 2 of Definition 3  holds for the states $\rho_{AB}$ as Eq. (\ref{eq:112}). Therefore,  Definitions 3 and $3^{\prime}$ are equivalent.

Let's end this section with examining the second property of  Definition 3, for a special interesting case. Consider a quadripartite SM state $\rho_{AE_{A}BE_{B}}$ . We, e.g.,  have for the CP map $\Lambda_{A}$
\begin{equation}
\label{eq:114}
\begin{aligned}
 \Lambda_{A}\otimes id_{B}(\rho_{AB})=\Lambda_{A}\otimes\Phi_{B}(\rho_{AB}) \\
 =\mathrm{Tr_{E_{B}}}\circ [\Lambda_{A}\otimes \Lambda_{B}](\rho_{AB})=\rho_{AE_{A}B}.
\end{aligned}
\end{equation}
So,
\begin{equation}
\label{eq:115}
\begin{aligned}
 id_{AE_{A}}\otimes \Lambda_{B}(\rho_{AE_{A}B})=\Lambda_{A}\otimes\Lambda_{B}(\rho_{AB})
=\rho_{AE_{A}BE_{B}};
\end{aligned}
\end{equation}
i.e., according to tripartition $(AE_{A};B;E_{B})$,
 $\rho_{AE_{A}BE_{B}}$ is a tripartite Markov state and can be written as
  Eq. (\ref{eq:4}). This, also, can be shown directly from Eq.   (\ref{eq:8}).

\section{The multipartite case}\label{sec:multi}

Now, we consider the case that the system is $N$-partite, $S=S_{1}S_{2}\dots S_{N}$. Different parts of the system are separated from 
each other and each part $S_{i}$ interacts with its  local environment $E_{i}$. 
We 
 denote the whole state of the system-environment as $\rho_{SE}=\rho_{S_{1}E_{1}\dots S_{N}E_{N}}$.

\textbf{Definition 4.} \textit{We call a $2N$-partite
 state $\rho_{S_{1}E_{1}\dots S_{N}E_{N}}$  a weak Markov (WM) state if there exist CP maps $\Lambda_{i}$, from $S_{i}$ to $S_{i}E_{i}$, such that}
\begin{equation}
\label{eq:120}
\begin{aligned}
\rho_{S_{1}E_{1}\dots S_{N}E_{N}}=\Lambda_{1}\otimes\Lambda_{2}\otimes \dots\otimes\Lambda_{N} (\rho_{S_{1}S_{2}\dots S_{N}}),  
\end{aligned}
\end{equation}
\textit{where $\rho_{S_{1}S_{2}\dots S_{N}}= \mathrm{Tr_{E_{1}\dots E_{N}}}(\rho_{S_{1}E_{1}\dots S_{N}E_{N}})$.}

Therefore, for a WM state, each localized dynamics as  $\mathcal{F}_{S_{1}E_{1}}\otimes\dots\otimes \mathcal{F}_{S_{N}E_{N}}$, for the whole system-environment, reduces to a localized subdynamics as $\mathcal{E}_{S_{1}}\otimes\dots\otimes \mathcal{E}_{S_{N}}$, for the system. So, we readily conclude that:

\textbf{Corollary 3.} \textit{If for a localized dynamics of the whole system-environment as $\mathcal{F}_{S_{1}E_{1}}\otimes\dots\otimes \mathcal{F}_{S_{N}E_{N}}$, the entanglement of the system $S=S_{1}S_{2}\dots S_{N}$ increases, then  we conclude that the initial state of the whole system-environment,  $\rho_{S_{1}E_{1}\dots S_{N}E_{N}}$, is not a WM state as Eq. (\ref{eq:120}).}

If we define the CP map $\Phi_{i}\equiv \mathrm{Tr_{E_{i}}}\circ\Lambda_{i}$ on the subsystem $S_{i}$, then, from Eq. (\ref{eq:120}), we have
\begin{equation}
\label{eq:121}
\begin{aligned}
\Phi_{1}\otimes\Phi_{2}\otimes \dots\otimes\Phi_{N} (\rho_{S_{1}S_{2}\dots S_{N}})=\rho_{S_{1}S_{2}\dots S_{N}}.  
\end{aligned}
\end{equation}

Now, as the previous section, we define a $2N$-partite SM state as the following:

\textbf{Definition 5.} \textit{We call a $2N$-partite
 state $\rho_{S_{1}E_{1}\dots S_{N}E_{N}}$  a strong Markov (SM) state if}
 
\textit{ 1.  Eq. (\ref{eq:120}) holds for it, and}
 
\textit{ 2. in Eq. (\ref{eq:121}), we can replace one or more $\Phi_{i}$ with $id_{S_{i}}$.}

\textbf{Theorem 4.} \textit{A $2N$-partite
 state $\rho_{S_{1}E_{1}\dots S_{N}E_{N}}$ is a strong Markov (SM) state, if and only if,  there exist  decompositions of the Hilbert spaces of the subsystems $S_{i}$, $\cH_{S_{i}}$, as $\cH_{S_{i}}=\bigoplus_{j_{i}}\cH_{(s_{i})_{j_{i}}^{L}}\otimes\cH_{(s_{i})_{j_{i}}^{R}}$,  such that} 
\begin{equation}
\label{eq:122}
\begin{aligned}
\rho_{S_{1}E_{1}\dots S_{N}E_{N}}
=\bigoplus_{j_{1}, \dots ,  j_{N}}\lambda_{j_{1}\dots j_{N}}\, \rho_{(s_{1})_{j_{1}}^{L}\dots(s_{N})_{j_{N}}^{L} }  \quad\qquad \\
\otimes\rho_{(s_{1})_{j_{1}}^{R}E_{1}}\otimes \dots\otimes \rho_{(s_{N})_{j_{N}}^{R}E_{N}}.
\end{aligned}
\end{equation}
\textit{In Eq.  (\ref{eq:122}),  $\lbrace \lambda_{j_{1}\dots j_{N}}\rbrace$  is a probability distribution, $\rho_{(s_{1})_{j_{1}}^{L}\dots(s_{N})_{j_{N}}^{L} }$ is a state on $\cH_{(s_{1})_{j_{1}}^{L}}\otimes\dots\otimes\cH_{(s_{N})_{j_{N}}^{L}}$  and $\rho_{(s_{i})_{j_{i}}^{R}E_{i}}$ is a state on $\cH_{(s_{i})_{j_{i}}^{R}}\otimes\cH_{E_{i}}$. ($ \cH_{E_{i}}$  is the Hilbert space of $E_{i}$.) }

\textit{Proof.} Showing that a $2N$-partite
 state  as Eq. (\ref{eq:122}) is a SM state, as Definition 5, is not difficult. It can be done by noting that, for a state as  (\ref{eq:122}), we have
\begin{equation}
\label{eq:123}
\begin{aligned}
\Lambda_{i}=\bigoplus_{j_{i}}id_{(s_{i})_{j_{i}}^{L}}\otimes \Lambda_{(s_{i})_{j_{i}}^{R}},
\end{aligned}
\end{equation}
where $id_{(s_{i})_{j_{i}}^{L}}$ is the identity map on $(s_{i})_{j_{i}}^{L}$ and $\Lambda_{(s_{i})_{j_{i}}^{R}}$ is a CP map from $(s_{i})_{j_{i}}^{R}$ to $(s_{i})_{j_{i}}^{R}E_{i}$.

So, we focus on proving the reverse: Each $2N$-partite SM state, as Definition 5, can be decomposed as  Eq. (\ref{eq:122}).

From the property 2 of  Definition 5, we know that, according to the bipartition $S_{1};S_{2}\dots S_{N}$, we have
\begin{equation*}
\label{eq:124}
\begin{aligned}
\Phi_{1}\otimes id_{S_{2}\dots S_{N}}(\rho_{S_{1};S_{2}\dots S_{N}})=\rho_{S_{1};S_{2}\dots S_{N}}.
\end{aligned}
\end{equation*}
It is similar to Eq. (\ref{eq:108}). So, we conclude that there exists  
a decomposition of the $\cH_{S_{1}}$ as $\cH_{S_{1}}=\bigoplus_{j_{1}}\cH_{(s_{1})^{L}_{j_{1}}}\otimes\cH_{(s_{1})^{R}_{j_{1}}}$  such that 
$\rho_{S_{1}\dots S_{N}}$ can be decomposed as
\begin{equation}
\label{eq:125}
\rho_{S_{1};S_{2}\dots S_{N}}=\bigoplus_{j_{1}}q_{j_{1}}\:\rho_{(s_{1})^{L}_{j_{1}} S_{2}\dots S_{N}}\otimes\rho_{(s_{1})^{R}_{j_{1}}},
\end{equation}
where $\lbrace q_{j_{1}}\rbrace$ is a probability distribution,  $\rho_{(s_{1})^{L}_{j_{1}}S_{2}\dots S_{N}}$ is a state on $\cH_{(s_{1})^{L}_{j_{1}}}\otimes\cH_{S_{2}\dots S_{N}}$ and $\rho_{(s_{1})^{R}_{j_{1}}}$ is a state on $\cH_{(s_{1})^{R}_{j_{1}}}$.

Similarly, according to the bipartition $S_{2};S_{1}S_{3}\dots S_{N}$, we have $\Phi_{2}\otimes id_{S_{1}S_{3}\dots S_{N}}(\rho_{S_{1}\dots S_{N}})=\rho_{S_{1}\dots S_{N}}$ and so
\begin{equation}
\label{eq:126}
\begin{aligned}
\rho_{S_{1}\dots S_{N}}=\bigoplus_{j_{2}}q_{j_{2}}\:\rho_{S_{1}(s_{2})^{L}_{j_{2}} S_{3}\dots S_{N}}\otimes\rho_{(s_{2})^{R}_{j_{2}}}, \\
\cH_{S_{2}}=\bigoplus_{j_{2}}\cH_{(s_{2})^{L}_{j_{2}}}\otimes\cH_{(s_{2})^{R}_{j_{2}}}. \qquad\qquad
\end{aligned}
\end{equation}

By defining the projector $\Pi_{j_{1}}=\Pi_{(s_{1})_{j_{1}}^{L}}\otimes \Pi_{(s_{1})_{j_{1}}^{R}}\otimes I_{S_{2}\dots S_{N}}$, from Eqs. (\ref{eq:125}) and (\ref{eq:126}),  we have
\begin{equation*}
\label{eq:127}
\begin{aligned}
\Pi_{j_{1}}\rho_{S_{1}\dots S_{N}}\Pi_{j_{1}}=q_{j_{1}}\:\rho_{(s_{1})^{L}_{j_{1}} S_{2}\dots S_{N}}\otimes\rho_{(s_{1})^{R}_{j_{1}}} \\
\bigoplus_{j_{2}}q_{j_{2}}\:\sigma_{(S_{1})_{j_{1}}(s_{2})^{L}_{j_{2}} S_{3}\dots S_{N}}\otimes\rho_{(s_{2})^{R}_{j_{2}}}, 
\end{aligned}
\end{equation*}
where $\sigma_{(S_{1})_{j_{1}}(s_{2})^{L}_{j_{2}} S_{3}\dots S_{N}}=\bar{\Pi}_{j_{1}} \rho_{ S_{1}(s_{2})^{L}_{j_{2}}S_{3}\dots S_{N}} \bar{\Pi}_{j_{1}}$, with $\bar{\Pi}_{j_{1}}=\Pi_{(s_{1})_{j_{1}}^{L}}\otimes \Pi_{(s_{1})_{j_{1}}^{R}}\otimes \Pi_{(s_{2})_{j_{2}}^{L}}\otimes I_{S_{3}\dots S_{N}}$. So, by a similar line of reasoning, as obtained from Eqs. (\ref{eq:111})-(\ref{eq:112}), we achieve

\begin{equation}
\label{eq:128}
\begin{aligned}
\rho_{S_{1}\dots S_{N}}=\bigoplus_{j_{1}, j_{2}}
\lambda_{j_{1}j_{2}}\:\rho_{(s_{1})^{L}_{j_{1}}(s_{2})^{L}_{j_{2}} S_{3}\dots S_{N}}
\otimes\rho_{(s_{1})^{R}_{j_{1}}}\otimes\rho_{(s_{2})^{R}_{j_{2}}}.
\end{aligned}
\end{equation}
By continuing this method, we finally get 
\begin{equation}
\label{eq:129}
\begin{aligned}
\rho_{S_{1}\dots S_{N}}
=\bigoplus_{j_{1}, \dots ,  j_{N}}\lambda_{j_{1}\dots j_{N}}\, \rho_{(s_{1})_{j_{1}}^{L}\dots(s_{N})_{j_{N}}^{L} }  \quad\qquad \\
\otimes\rho_{(s_{1})_{j_{1}}^{R}}\otimes \dots\otimes \rho_{(s_{N})_{j_{N}}^{R}}, \\
\cH_{S_{i}}=\bigoplus_{j_{i}}\cH_{(s_{i})_{j_{i}}^{L}}\otimes\cH_{(s_{i})_{j_{i}}^{R}}.  \qquad\qquad
\end{aligned}
\end{equation}
In Eq.  (\ref{eq:129}),  $\lbrace \lambda_{j_{1}\dots j_{N}}\rbrace$  is a probability distribution, $\rho_{(s_{1})_{j_{1}}^{L}\dots(s_{N})_{j_{N}}^{L} }$ is a state on $\cH_{(s_{1})_{j_{1}}^{L}}\otimes\dots\otimes\cH_{(s_{N})_{j_{N}}^{L}}$  and $\rho_{(s_{i})_{j_{i}}^{R}}$ is a state on $\cH_{(s_{i})_{j_{i}}^{R}}$. 

Next, note that, during the proof of Theorem 3, from Eq. (\ref{eq:108}), we have concluded Eq. (\ref{eq:seventeen}). Here also, from 
\begin{equation*}
\label{eq:130}
\begin{aligned}
\Phi_{i}\otimes id_{S_{1}\dots S_{i-1}S_{i+1}\dots S_{N}}(\rho_{S_{1}\dots S_{N}})=\rho_{S_{1}\dots S_{N}},
\end{aligned}
\end{equation*}
we conclude Eq. (\ref{eq:123}). So, from Eqs. (\ref{eq:120}), (\ref{eq:123}) and (\ref{eq:129}), we achieve Eq. (\ref{eq:122}), and the proof is completed.
$\qquad\square$


\textbf{Remark 3.} \textit{Theorem 4 is valid for the case that $\cH_{S_{i}}$ are finite dimensional, but $\cH_{E_{i}}$  can be infinite dimensional. In other words, the system $S=S_{1}\dots S_{N}$ is finite dimensional but the environments  $E_{i}$  can be infinite dimensional.}

As stated in Corollary 3, for a WM state, each localized dynamics as 
$\mathcal{F}_{S_{1}E_{1}}\otimes\dots\otimes \mathcal{F}_{S_{N}E_{N}}$ reduces to a localized subdynamics as $\mathcal{E}_{S_{1}}\otimes\dots\otimes \mathcal{E}_{S_{N}}$. Now, for an SM state, from Eqs. (\ref{eq:123}) and (\ref{eq:129}), it can be shown that if $\mathcal{F}_{S_{i}E_{i}}=id_{S_{i}E_{i}}$, then $\mathcal{E}_{S_{i}}=id_{S_{i}}$.

Till now, we have considered the case that our $M$-partite Markov state includes even subsystems: $M=2N$. In the following, we consider the case that $M=2N-1$, $N=3, 4, \dots$. The case that $N=2$, and so $M=3$, has been considered in Sect.~\ref{sec:part B}.

 Consider the case that the system is $N$-partite, $S=S_{1}S_{2}\dots S_{N}$. The part $S_{1}$ is isolated and the other parts
    $S_{i}$, $i\neq 1$, each interacts with its  local environment $E_{i}$. 
We 
 denote the whole state of the system-environment as $\rho_{SE}=\rho_{S_{1}S_{2}E_{2}\dots S_{N}E_{N}}$.

\textbf{Definition 6.} \textit{We call a $(2N-1)$-partite
 state $\rho_{S_{1}S_{2}E_{2}\dots S_{N}E_{N}}$  a strong Markov (SM) state if}
 
\textit{ 1.  there exist CP maps $\Lambda_{i}$, from $S_{i}$ to $S_{i}E_{i}$, $i\neq 1$, such that}
\begin{equation}
\label{eq:131}
\begin{aligned}
\rho_{S_{1}S_{2}E_{2}\dots S_{N}E_{N}}=id_{S_{1}}\otimes\Lambda_{2}\otimes \dots\otimes\Lambda_{N} (\rho_{S_{1}S_{2}\dots S_{N}}),  
\end{aligned}
\end{equation}
\textit{where $\rho_{S_{1}S_{2}\dots S_{N}}= \mathrm{Tr_{E_{2}\dots E_{N}}}(\rho_{S_{1}S_{2}E_{2}\dots S_{N}E_{N}})$, and}

\textit{2 in the relation}
 \begin{equation}
\label{eq:132}
\begin{aligned}
id_{S_{1}}\otimes\Phi_{2}\otimes \dots\otimes\Phi_{N} (\rho_{S_{1}S_{2}\dots S_{N}})=\rho_{S_{1}S_{2}\dots S_{N}},  
\end{aligned}
\end{equation}
\textit{where  $\Phi_{i}\equiv \mathrm{Tr_{E_{i}}}\circ\Lambda_{i}$, $i\neq 1$, is a CP map on $S_{i}$,  we can replace one or more  $\Phi_{i}$ with $id_{S_{i}}$.}

In addition, we call a $(2N-1)$-partite
 state $\rho_{S_{1}S_{2}E_{2}\dots S_{N}E_{N}}$   a weak Markov (WM) state, if it only possesses the property 1, in the above definition. Obviously, for a WM state, each localized dynamics as $id_{S_{1}}\otimes 
\mathcal{F}_{S_{2}E_{2}}\otimes\dots\otimes \mathcal{F}_{S_{N}E_{N}}$, for the whole system-environment, reduces to a localized subdynamics as $id_{S_{1}}\otimes\mathcal{E}_{S_{2}}\otimes\dots\otimes \mathcal{E}_{S_{N}}$, for the system. Therefore, a result, similar to  Corollary 3, can be obtained for this case, too.

In the following of this section, we give our final main result: The structure of the $(2N-1)$-partite SM states.

\textbf{Theorem 5.} \textit{A $(2N-1)$-partite
 state $\rho_{S_{1}S_{2}E_{2}\dots S_{N}E_{N}}$ is a strong Markov (SM) state, if and only if,  there exist  decompositions of the Hilbert spaces of the subsystems $S_{i}$, $\cH_{S_{i}}$, $i\neq 1$, as $\cH_{S_{i}}=\bigoplus_{j_{i}}\cH_{(s_{i})_{j_{i}}^{L}}\otimes\cH_{(s_{i})_{j_{i}}^{R}}$,  such that} 
\begin{equation}
\label{eq:133}
\begin{aligned}
\rho_{S_{1}S_{2}E_{2}\dots S_{N}E_{N}}
=\bigoplus_{j_{2}, \dots ,  j_{N}}\lambda_{j_{2}\dots j_{N}}\, \rho_{S_{1}(s_{2})_{j_{2}}^{L}\dots(s_{N})_{j_{N}}^{L} }  \quad\qquad \\
\otimes\rho_{(s_{2})_{j_{2}}^{R}E_{2}}\otimes \dots\otimes \rho_{(s_{N})_{j_{N}}^{R}E_{N}}.
\end{aligned}
\end{equation}
\textit{In Eq.  (\ref{eq:133}),  $\lbrace \lambda_{j_{2}\dots j_{N}}\rbrace$  is a probability distribution, $\rho_{S_{1}(s_{2})_{j_{2}}^{L}\dots(s_{N})_{j_{N}}^{L} }$ is a state on $\cH_{S_{1}}\otimes\cH_{(s_{2})_{j_{2}}^{L}}\otimes\dots\otimes\cH_{(s_{N})_{j_{N}}^{L}}$  and $\rho_{(s_{i})_{j_{i}}^{R}E_{i}}$ is a state on $\cH_{(s_{i})_{j_{i}}^{R}}\otimes\cH_{E_{i}}$.}

\textit{Proof.} As  Theorem 4, proving that a state given in Eq. (\ref{eq:133}) is a SM state, as the Definition 6, is not difficult, since, here also, the CP maps $\Lambda_{i}$ are as Eq. (\ref{eq:123}).

The proof of the reverse, i.e. each $(2N-1)$-partite SM
 state can be decomposed as Eq. (\ref{eq:133}), is also similar to what has been done during the proof of Theorem 4.
The only difference is that the starting point is the Eq. (\ref{eq:126}), instead of Eq. (\ref{eq:125}). So, instead of Eq. (\ref{eq:129}), we achieve
\begin{equation}
\label{eq:134}
\begin{aligned}
\rho_{S_{1}\dots S_{N}}
=\bigoplus_{j_{2}, \dots ,  j_{N}}\lambda_{j_{2}\dots j_{N}}\, \rho_{S_{1}(s_{2})_{j_{2}}^{L}\dots(s_{N})_{j_{N}}^{L} }  \quad\qquad \\
\otimes\rho_{(s_{2})_{j_{2}}^{R}}\otimes \dots\otimes \rho_{(s_{N})_{j_{N}}^{R}}, \\
\cH_{S_{i}}=\bigoplus_{j_{i}}\cH_{(s_{i})_{j_{i}}^{L}}\otimes\cH_{(s_{i})_{j_{i}}^{R}} \qquad (i\neq 1).  \qquad
\end{aligned}
\end{equation}
In Eq.  (\ref{eq:134}),  $\lbrace \lambda_{j_{2}\dots j_{N}}\rbrace$  is a probability distribution, $\rho_{S_{1}(s_{2})_{j_{2}}^{L}\dots(s_{N})_{j_{N}}^{L} }$ is a state on $\cH_{S_{1}}\otimes\cH_{(s_{2})_{j_{2}}^{L}}\otimes\dots\otimes\cH_{(s_{N})_{j_{N}}^{L}}$  and $\rho_{(s_{i})_{j_{i}}^{R}}$ is a state on $\cH_{(s_{i})_{j_{i}}^{R}}$. 

Then, using Eqs. (\ref{eq:123}), (\ref{eq:131}) and (\ref{eq:134}), we get Eq. (\ref{eq:133}), and the proof is completed. 
$\quad\qquad\qquad\qquad\qquad\qquad\square$

\textbf{Remark 4.} \textit{Theorem 5 is valid for the case that $\cH_{S_{i}}$, $i\neq 1$, are finite dimensional, but $\cH_{S_{1}}$ and $\cH_{E_{i}}$  can be infinite dimensional.}

Till now, we have defined the Markov states for all $M$-partite cases for which $M=3, 4, \dots$. The generalization to the cases that $M=1, 2$ is straightforward and may be interesting. So, we give them in the following.

We can call each one-partite state $\rho_{S}$ a Markov state, since there is a CP map, i.e., $id_{S}$, such that $\rho_{S}=id_{S}(\rho_{S})$.

In addition, we can call each bipartite state $\rho_{SE}$ a Markov state, too. It is so since one can find  a CP map $\Lambda$, from $S$ to $SE$, such that $\rho_{SE}=\Lambda(\rho_{S})$, where $\rho_{S}=\mathrm{Tr_{E}}(\rho_{SE})$.
For example, $\Lambda$ can be constructed as $\Lambda=\bar{\Lambda} \circ \Xi$. The CP map $\Xi$ is defined as $\Xi(\rho_{S})=(I_{S}\otimes\vert0_{E}\rangle)\,\rho_{S}\,(I_{S}\otimes\langle 0_{E}\vert)$, where $\vert0_{E}\rangle$ is a fixed state in $\cH_{E}$. The completely positive map $\bar{\Lambda}$, which maps $\rho_{S}\otimes\vert0_{E}\rangle\langle0_{E}\vert$ to the $\rho_{SE}$, can be found, e.g., using the method introduced in Ref. \cite{19}.

\section{Example: the classical environment}\label{sec:classical}

We end our paper with an example of the simplest case, i.e., the case studied in Sect. ~\ref{sec:part B}.          Some other examples  are also given in Ref. \cite{42}.

Consider the case that the system $S$ is bipartite, $S=AB$. The part $A$ is isolated from the environment  and only the part $B$ interacts with the environment $E$.
 In addition, assume that the effect of $E$ on $B$ can be modeled as acting random unitary operators $U^{(j)}_{B}$ on $B$, each with the probability $p_{j}$. Therefore, the whole dynamics of the system can be written as

\begin{equation}
\label{eq:1}
\begin{aligned}
\rho_{AB}(t)=\sum_{j}p_{j}\, \left( I_{A}\otimes U^{(j)}_{B}(t)\right)\,\rho_{AB}(0)\, \left( I_{A}\otimes U^{(j)\dagger}_{B}(t)\right),
\end{aligned}
\end{equation}
where $I_{A}$ is the identity operator on $A$, $\rho_{AB}(0)$ is the initial state of the system and $\rho_{AB}(t)$ is the state of the system at time $t$. In  Eq. (\ref{eq:1}), $U^{(j)}_{B}(t)=U^{(j)}_{B}(t,0)$ is a unitary time evolution, acting on $B$ with the probability $p_{j}$, from the initial moment to the time  $t$.
 Note that $U^{(j)}_{B}(t_{2},0)=U^{(j)}_{B}(t_{2},t_{1})U^{(j)}_{B}(t_{1},0)$. In the simplest case, we have  $U^{(j)}_{B}(t)=e^{-iH_{j}t/\hbar}$, with a  time-independent Hamiltonian $H_{j}$.
 
 In Refs. \cite{5, 6, 7, 8}, some quantum systems, for which the time evolution is given by  Eq. (\ref{eq:1}), are studied. An important example is when the subsystem $B$ is coupled to a random external field and the subsystem $A$ is isolated from this classical external field \cite{5, 6}. The characteristics of the classical external field are not affected by interaction with the $B$ and so its state remains unchanged during the evolution.
 
 We can model the whole system-environment evolution as the following \cite{4}. We get the initial state of the system-environment as
 \begin{equation}
\label{eq:2}
\begin{aligned}
\rho_{SE}(0)=\rho_{AB}(0)\otimes\sum_{j}p_{j}\vert j_{E}\rangle\langle j_{E}\vert ,
\end{aligned}
\end{equation} 
where $\lbrace\vert j_{E}\rangle\rbrace$ is an orthonormal basis for $E$. In addition, the system-environment undergoes the evolution given by the unitary operator
 \begin{equation}
\label{eq:3}
\begin{aligned}
U_{SE}(t)=\sum_{j} I_{A}\otimes U^{(j)}_{B}(t) \otimes\vert j_{E}\rangle\langle j_{E}\vert .
\end{aligned}
\end{equation}
From Eqs. (\ref{eq:2}) and (\ref{eq:3}), it can be shown simply that the reduced dynamics of the system $S=AB$ is given by Eq. (\ref{eq:1}). In addition, the reduced state of the environment remains unchanged during the evolution. We have $\rho_{E}(t)=\sum_{j}p_{j}\vert j_{E}\rangle\langle j_{E}\vert=\rho_{E}(0)$, which is a classical state, i.e., it contains no superposition of the basis states $\vert j_{E}\rangle$. 

If the initial state of the system, $\rho_{AB}(0)$, be an entangled state, since the environment is classical, we may expect that, during the time evolution of the system, entanglement decreases monotonically. But, unexpectedly, it has been shown, both theoretically and experimentally, that for a system which undergoes the  time evolution given by Eq. (\ref{eq:1}), entanglement revivals can occur \cite{3, 5, 6, 7, 8}.



Note that, $\rho_{ABE}(0)$ in Eq. (\ref{eq:2}) is a Markov state; that is, it can be written in the form of  Eq. (\ref{eq:4}). It is, in fact, a factorized state which is due to the case that $\cH_{B}=\cH_{b^{L}}\otimes\cH_{b^{R}}$ and $\cH_{b^{R}}$ is a trivial one-dimensional Hilbert space. In addition, the dynamics of the system-environment in Eq. (\ref{eq:3}) is localized as  Eq. (\ref{eq:101}). Therefore, the reduced dynamics of the system in Eq. (\ref{eq:1}) is also localized as Eq. (\ref{eq:103}). So, $\mathcal{M}(\rho_{AB}(t))\leq \mathcal{M}(\rho_{AB}(0))$, for all $t>0$. This is in agreement with the results of Refs. \cite{3, 5, 6, 7, 8}.

From Eq. (\ref{eq:3}), we see that the time evolution operator of the system-environment, from $t_{1}$ to $t_{2}$ $(t_{1} < t_{2})$, is as 
\begin{equation}
\label{eq:7}
\begin{aligned}
U_{ABE}(t_{2},t_{1})=I_{A}\otimes \left(\sum_{j}  U^{(j)}_{B}(t_{2},t_{1}) \otimes\vert j_{E}\rangle\langle j_{E}\vert\right),
\end{aligned}
\end{equation}
which is in the form of Eq. (\ref{eq:101}). Therefore, if, at time $t=t_{1}$, entanglement starts to increase, it indicates that $\rho_{ABE}(t_{1})$ is not a Markov state.  Note that the state of this hybrid quantum-classical system $SE$ changes from its initial factorized state $\rho_{ABE}(0)$ in Eq.  (\ref{eq:2}) to the state $\rho_{ABE}(t_{1})$, which cannot be written as Eq.  (\ref{eq:4}). So, although the reduced state of $E$ remains unchanged during the evolution,  the whole state of the system-environment changes from its initial factorized one to a state which is not a Markov state and this change can lead to the entanglement revival.

As we see in the following, during the evolution, though  the whole state of the system-environment changes from its initial Markov state to a non-Markovian state, but the correlation between the system $S=AB$ and the environment $E$ remains classical. This implies that the non-Markovianity of the $\rho_{SE}$ is not equivalent to existence of non-classical correlation between $S$ and $E$.

If we define the one-dimensional projectors $\Pi_{E}^{(j)}=\vert j_{E}\rangle\langle j_{E}\vert$, then,
 from Eq. (\ref{eq:2}), it can be seen that $\rho_{SE}(0)$ does not change under local projective measurement $\lbrace I_{S}\otimes \Pi_{E}^{(j)}\rbrace$; that is, if we perform the measurement $\lbrace I_{S}\otimes \Pi_{E}^{(j)}\rbrace$ on $\rho_{SE}(0)$ and then mix the results of different outcomes, we achieve the pre-measurement state $\rho_{SE}(0)$.
This can be interpreted as the existence of no quantum correlation between $S$ and $E$ \cite{20}.

The above argument is also true for  $\rho_{SE}(t)$. From Eqs. (\ref{eq:2}) and (\ref{eq:3}), we have
 \begin{equation}
\label{eq:333}
\begin{aligned}
\rho_{SE}(t)=\sum_{j}p_{j} \rho_{AB}^{(j)}(t)\otimes\vert j_{E}\rangle\langle j_{E}\vert,
\end{aligned}
\end{equation}
where 
\begin{equation}
\label{eq:334}
\rho_{AB}^{(j)}(t)=I_{A}\otimes U^{(j)}_{B}\rho_{AB}(0)I_{A}\otimes U^{(j)\dagger}_{B}.
\end{equation}
So, $\rho_{SE}(t)$ is also unchanged under the measurement  $\lbrace I_{S}\otimes \Pi_{E}^{(j)}\rbrace$ and the correlation between the system $S=AB$ and the environment $E$ remains classical during the evolution.

Note that, in Eq. (\ref{eq:333}), each $\rho_{AB}^{(j)}(t)$ is coupled to a fixed unchanged state of the environment $\vert j_{E}\rangle\langle j_{E}\vert$. As expected, there is no correlation between the  $\rho_{AB}^{(j)}(t)$ and $\vert j_{E}\rangle\langle j_{E}\vert$, since the environment is classical and unchanged during the evolution. So, any classical correlation in Eq. (\ref{eq:333}) is due to the mixing different $\rho_{AB}^{(j)}(t)\otimes\vert j_{E}\rangle\langle j_{E}\vert$, each with the probability $p_{j}$. 

In fact, we are encountered with an ensemble of the states as $\lbrace p_{j},  \rho_{AB}^{(j)}(t)\otimes\vert j_{E}\rangle\langle j_{E}\vert \rbrace$.
Therefore, it can be argued \cite{7} that the real amount of entanglement present between $A$ and $B$ is
 \begin{equation}
\label{eq:335}
\begin{aligned}
\sum_{j}p_{j} \mathcal{M} (\rho_{AB}^{(j)}(t)) \qquad\qquad \\
=\sum_{j}p_{j} \mathcal{M} (\rho_{AB}(0))=\mathcal{M} (\rho_{AB}(0)),
\end{aligned}
\end{equation}
where we have used this fact that under local operation, in Eq. (\ref{eq:334}), entanglement between $A$ and $B$ does not change.
The only reason which prevent us to achieve all of this amount is the mixing in Eq. (\ref{eq:333}). So, one can define the \textit{hidden entanglement} as \cite{7}:
 \begin{equation}
\label{eq:336}
\begin{aligned}
\mathcal{M}_{H}(t)=\sum_{j}p_{j} \mathcal{M} (\rho_{AB}^{(j)}(t)) - \mathcal{M}(\rho_{AB}(t)) \\
=\mathcal{M} (\rho_{AB}(0))- \mathcal{M}(\rho_{AB}(t)), \qquad
\end{aligned}
\end{equation}
which gives the amount of entanglement, though present between $A$ and $B$, is hidden (inaccessible) for us (see also Ref. \cite{42}).

\section{Summary }\label{sec:summary}

Markov states has been defined for the tripartite case \cite{10}.
In this paper, we have generalized the definition of the Markov state to arbitrary $M$-partite case.

We have given two forms of definitions: \textit{weak Markov} (WM) states and \textit{strong Markov} (SM) states. The set of SM states is a subset of the set of WM states. For $M\leq 3$, the two sets are the same. For $M>3$, though it seems that the set of SM states is a proper subset of the set of WM states, a careful treatment is needed to prove  
 whether these two sets are the same or not.

For WM states, we have seen that each localized dynamics for the whole system-environment reduces to a localized subdynamics of the system. This provides us a necessary (but, in general, insufficient) condition, for entanglement increase: 
Entanglement revival can occur only when the initial state of the system-environment state is not a WM state.

Our main results, in this paper, are for SM states. We have found the general structure of the SM states, for arbitrary $M$-partite case, in Theorems 3, 4 and 5.

 If the initial state of the whole system-environment  $\rho_{S_{1}E_{1}\dots S_{N}E_{N}}$ is a SM
 sate, then, since each SM state is, in addition, a WM state, each localized dynamics for the  system-environment as $\mathcal{F}_{S_{1}E_{1}}\otimes\dots\otimes \mathcal{F}_{S_{N}E_{N}}$ reduces to a localized subdynamics as $\mathcal{E}_{S_{1}}\otimes\dots\otimes \mathcal{E}_{S_{N}}$ for the system. Also,  if $\mathcal{F}_{S_{i}E_{i}}=id_{S_{i}E_{i}}$, then $\mathcal{E}_{S_{i}}=id_{S_{i}}$.
 
 According to the two above interesting properties, it seems that the SM states can play an important role in studying open quantum systems.
 
 We have ended our paper by studying an example of the simplest case, i.e., the tripartite case $\rho_{ABE}$. We have considered the case that though the environment $E$ is classical, entanglement revival can occur in the system $S=AB$. Entanglement revival can occur only when the whole state of the  
system-environment changes from its initial Markov state to a non-Markovian state. But, during this change, the correlation between the system $S$ and the environment $E$ remains classical. This implies that the non-Markovianity of a state is not equivalent to existence of non-classical correlation between the system and the environment.

\end{document}